\documentclass[10pt,letterpaper]{article}
\usepackage{opex3}

\begin{document}
\title{Spatial coherence effects on second- and fourth-order temporal interference}
\author{Timothy Yarnall$^{1,2}$, Ayman F. Abouraddy$^{3}$, Bahaa E. A. Saleh$^{2}$, and Malvin C. Teich$^{2}$}
\address{$^{1}$ Lincoln Laboratory, Massachusetts Institute of Technology, 244 Wood Street, Lexington, Massachusetts, 02420-9108, USA}
\vspace{-10pt}
\address{$^{2}$Quantum Imaging Laboratory, Departments of Electrical
\& Computer Engineering and Physics, Boston University, Boston,
Massachusetts 02215-2421, USA}
\vspace{-10pt}
\address{$^{3}$Research Laboratory of Electronics, Massachusetts
Institute of Technology, Cambridge, Massachusetts 02139-4307, USA}

\date{\today}

\begin{abstract}
We report the results of two experiments performed with two-photon
light, produced via collinear degenerate optical spontaneous
parametric downconversion (SPDC), in which both second-order
(one-photon) and fourth-order (two-photon) interferograms are
recorded in a Mach--Zehnder interferometer (MZI).  In the first
experiment, high-visibility fringes are obtained for both the
second- and fourth-order interferograms. In the second experiment,
the MZI is modified by the removal of a mirror from one of its arms;
this leaves the fourth-order interferogram unchanged, but
extinguishes the second-order interferogram. A theoretical model
that takes into consideration both the temporal and spatial
degrees-of-freedom of the two-photon state successfully explains the
results. While the temporal interference in the MZI is independent
of the spatial coherence of the source, that of the modified MZI is
not.
\end{abstract}


\section{Introduction}
Two-photon interference experiments reflect correlations that are
fourth-order in the optical field and second-order in the intensity,
while one-photon interference experiments reflect correlations that
are second-order in the field and first-order in the intensity. It
is sometimes taken as a rule-of-thumb in quantum optics that
two-photon interference experiments displaying high two-photon
visibility $V_{12}$ must necessarily display low one-photon
visibility $V_{1}$ \cite{Horne89PRL}. We examine this issue
experimentally by carrying out two-photon (coincidence) measurements
using a pair of detectors to determine the correlation function
$G^{(2)}(\tau)$ and, simultaneously, one-photon (singles)
measurements using a single detector to determine the intensity
$I(\tau)$, for the same optical field.

We demonstrate in this paper that this rule-of-thumb can be
rigorously supported only when each photon of the two-photon state
is sent into a separate two-path interferometer that acts on a
single physical degree of freedom (and may thus be represented by an
SU(2) transformation) \cite{Greenberger93PT,Horne90Nature}. The
visibilities for a two-photon state in such a configuration are
indeed complementary such that they satisfy
$V_{12}^{2}+V_{1}^{2}\leq1$, where the equality holds for pure
two-photon states \cite{Jaeger93PRA,Jaeger95PRA}. This
complementarity is not applicable, however, when both photons are
directed into the \textit{same} interferometer, nor does it apply
when more than one degree-of-freedom is probed by the
interferometer. The most notable counterexamples of this
rule-of-thumb in temporal interferometry can be found in the
interferograms recorded with Hong--Ou--Mandel (HOM) \cite{HOM87PRL}
and Mach--Zehnder interferometers (MZI).

At first blush, experiments performed using a HOM interferometer in
conjunction with the two-photon state produced by spontaneous
parametric downconversion (SPDC) appear to confirm the
rule-of-thumb: when the temporal delay is scanned, high-visibility
two-photon interference (the HOM dip) is observed and $V_{12}=1$,
while the singles rate remains flat at $V_{1}=0$ \cite{HOM87PRL}. It
has recently been shown however, that by manipulating the
\textit{spatial} distribution of the pump, the visibility of the
temporal HOM dip may be varied from unity to zero, while the singles
rate remains unchanged (i.e., exhibiting no interference)
\cite{Walborn03PRL}. Moreover, it has been shown experimentally that
the rule-of-thumb does not hold for an MZI in which each photon of a
two-photon SPDC state was directed to a different input port of the
same MZI; high $V_{1}$ \textit{and} $V_{12}$ have been
simultaneously observed \cite{Larchuk93PRL}.

Clearly, complementarity in the visibilities of one- and two-photon
interference is not universally applicable. Other examples of
two-photon interferometry that make use of two-photon sources, in
which each photon is sent into a different interferometer, include
the experiments reported in
Refs.~\cite{Franson91PRA,Rarity92PRA,Strekalov96PRARC}. Examples in
which both photons are sent into the same interferometer include the
experiments reported in
Refs.~\cite{Rarity90PRL,Abouraddy01PRA,Santori02Nature}.

In this paper, we consider a slightly different question: can the
visibility of second-order interference be changed without affecting
the fourth-order interference? To address this issue, we examine the
two-photon interference in a MZI from a fresh perspective and show
that, surprisingly, \textit{high or low second-order} interference
($V_{1}=1$ or $V_{1}=0$) may be observed while retaining
\textit{high fourth-order} interference ($V_{12}=1$). This effect is
the opposite of that reported by Walborn \emph{et~al.}
\cite{Walborn03PRL}; moreover, in our case, the change in the
second-order interference visibility is achieved \textit{without}
altering the spatial distribution of the source. Our results thus
demonstrate both a confirmation of, and a departure from, the
rule-of-thumb in two closely related, but distinct, experimental
configurations.

\section{Experiment}
\begin{figure}[ht]
\centering\includegraphics[width=3.0in]{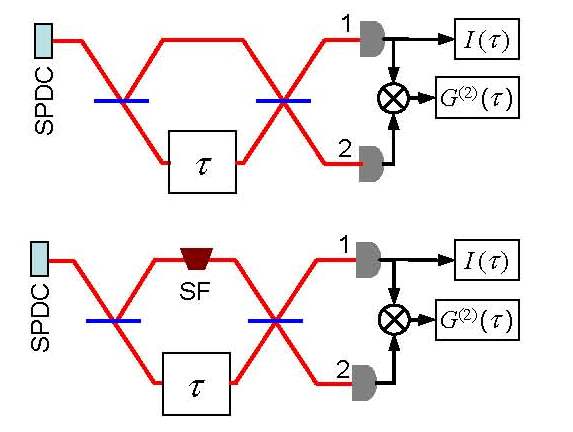}\\
  \caption{Conceptual layout for two
   Mach--Zehnder interferometer experiments. The upper diagram illustrates
   the traditional MZI while the lower diagram illustrates a modified version (MZIM) in which a
   spatial flip (SF) has been
   inserted into one arm. Both are followed
   by detectors that record the coincidence rate at the two output ports,
   $G^{(2)}(\tau)$, as well as the singles rate, $I(\tau)$.}
  \label{fig:Conceptual}
\end{figure}

The experimental configurations under discussion differ only in the
form of the Mach--Zehnder interferometer. As depicted in
Fig.~\ref{fig:Conceptual}, the first is a traditional MZI in which
the delay $\tau$ between the interfering paths is varied. The second
is a MZI in which a spatial flip (SF) has been inserted into one arm
of the interferometer
\cite{Sasada03PRA,Abouraddy07PRA,Yarnall07PRL1}; we refer to this as
a modified Mach-Zehnder interferometer (MZIM). The spatial flip is
achieved by unbalancing the number of mirrors in the two
interferometer arms; since the flip is carried out only in one
spatial dimension, it is readily implemented by simply removing or
adding a single mirror. No use is made of out-of-plane reflections,
such as those reported in Refs.~\cite{Segev92PRL,vanExter07OptExpr}.
In both cases, the input light is in a two-photon state produced via
SPDC \cite{Harris67PRL}, with both entangled photons entering the
same input port of the interferometer.

\begin{figure}[ht]
\centering\includegraphics[width=4.75in]{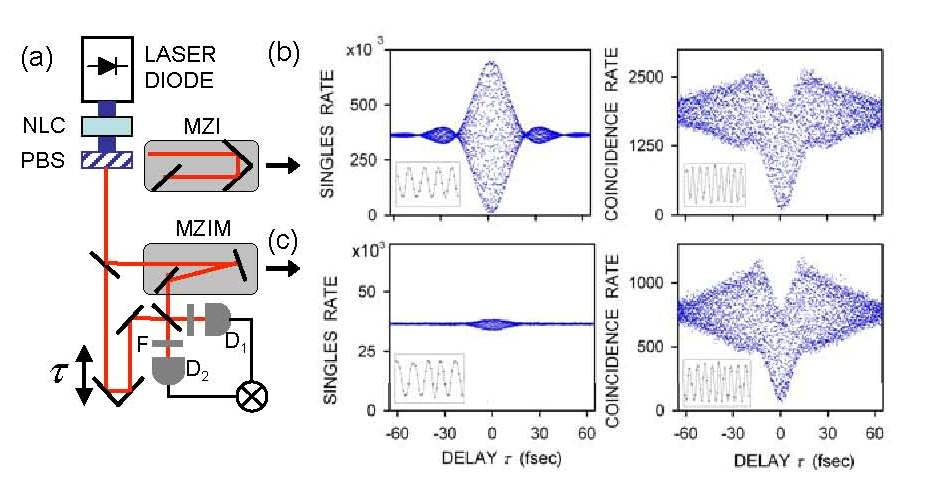}\\
\caption{(a) Schematic of the Mach-Zehnder experimental arrangements
for studying second- and fourth-order temporal interference. NLC:
nonlinear crystal; PBS: polarizing beam splitter; F: interference
filter; D: detector; $\otimes$: coincidence circuit. The upper
shaded region highlights one arm of the traditional Mach--Zehnder
interferometer (MZI), which comprises three mirrors, while the lower
shaded region highlights one arm of the modified Mach--Zehnder
interferometer (MZIM), which comprises two mirrors. (b) Singles and
coincidence interferograms for the MZI experiment. (c) Singles and
coincidence interferograms for the MZIM experiment. The coincidence
interferograms are similar while the singles interferograms are
distinctly different. The insets show data
in the vicinity of $\tau=0$ for each of the four plots; coincidence
interferograms oscillate at the pump period, with a frequency twice
that of the singles interferograms.}
  \label{fig:Setup and data}
\end{figure}

The setups for the MZI and MZIM experiments are shown in more detail
in Fig.~\ref{fig:Setup and data}(a). A linearly polarized
monochromatic pump laser diode (wavelength 405 nm, power 50 mW)
illuminates a 1.5-mm-thick $\beta$-barium borate (BBO) nonlinear
optical crystal (NLC) in a collinear type-I configuration (signal
and idler photons have the same polarization, orthogonal to that of
the pump). The pump is removed by using a polarizing beam splitter
placed after the crystal as well as by interference filters
(centered at 810 nm, 10-nm bandwidth) placed in front of the
detectors $\rm{D_1}$ and $\rm{D_2}$ (EG$\&$G SPCM-AQR-15-FC), the
outputs of which are fed to a coincidence circuit (denoted
$\otimes$) and thence to a counter. The bandwidth of the
interference filters is smaller than that of the SPDC-generated
two-photon state, so that the widths of temporal interference
features, such as the HOM dip and the MZI interferogram, are
expected to be proportional to the inverse of the filter bandwidth.

The singles and coincidence rates are presented in
Figs.~\ref{fig:Setup and data}(b) and ~\ref{fig:Setup and data}(c)
for the MZI and MZIM configurations, respectively. The MZI intensity
interferogram exhibits nearly $100\%$ visibility ($V_{1}=1$) and has
a sinc-like form consistent with a 10-nm bandpass filter of
approximately square spectral pass band. The fast oscillation occurs
at the down-converted photon frequency $\frac{\omega_{p}}{2}$, as
expected (see inset). The corresponding coincidence interferogram
exhibits fast oscillation at the pump period (see its inset)
together with an HOM dip. The MZIM coincidence interferogram is
nearly identical, but the intensity interferogram is essentially
flat, indicating the absence of second-order interference.

Our experimental results therefore reveal that the coincidence
measurements of photons at the two output ports yield essentially
identical outcomes for both interferometers: high-visibility
two-photon interference ($V_{12}=1$). Yet the singles measurements
yield opposite outcomes: the MZI reveals high-visibility one-photon
interference ($V_{1}=1$, in disagreement with the rule-of-thumb),
whereas the MZIM reveals the absence of one-photon interference
($V_{1} \approx 0$, in agreement with the rule-of-thumb). Unlike the
experiments carried out by Walborn et al. \cite{Walborn03PRL}, the
spatial distribution of the source was not modified in going from
one experiment to the other.

Let us consider the ways in which photon coincidences at the two
output ports of the interferometer may occur. From a simplified
viewpoint, there are two distinct possibilites that lead to
qualitatively different temporal interference features. In the first
of these, each photon emerges from a different port of the first
beam splitter. When these photons are brought back at the second
beam splitter, after a delay $\tau$ in one of the arms, an HOM dip
is observed in the coincidence rate $G^{(2)}(\tau)$. In the second
possibility, the two photons emerge together from either output port
of the first beam splitter. If the frequencies of the two photons,
signal and idler, are anti-correlated (which is the case for SPDC
with a monochromatic pump, so that
$\omega_s=\frac{\omega_{p}}{2}+\Omega$ and
$\omega_i=\frac{\omega_{p}}{2}-\Omega$, where $\frac{\omega_p}{2}$
is half the pump frequency and $\Omega$ is a deviation therefrom),
then a delay $\tau$ will then lead to a fixed phase difference
$\exp\{-i(\frac{\omega_{p}}{2}+\Omega)\tau\}\exp\{-i(\frac{\omega_{p}}{2}-\Omega)\tau\}=\exp\{-i\omega_{p}\tau\}$
between the two paths. In this case, $G^{(2)}(\tau)$ will be a
sinusoid at the pump period \cite{Rarity90PRL}. These two
possibilites coexist in the experimental arrangement shown in
Figs.~\ref{fig:Conceptual} and \ref{fig:Setup and data}(a),
resulting in a coincidence interferogram that combines an HOM dip
and a sinusoid at the pump period. Indeed, this is exactly what is
observed.

For the singles (intensity) rate at one output port, we expect the
usual MZI interferogram, proportional to the second-order temporal
coherence function of the optical field. The temporal width of this
interferogram should be inversely proportional to the bandwidth of
the optical field, as determined by either the source or the
detector filter, whichever is narrower. This is, in fact, true for
the MZI but, remarkably, it turns out \emph{not} to be true for the
MZIM. Rather, the loss of second-order \textit{temporal}
interference follows from the \textit{spatial coherence} properties
of the source, as we demonstrate below. This surprising result can
be understood when the full quantum state, including the spatial
distribution of the photons, is taken into consideration
\cite{Zeilinger98PS}.

\section{Theoretical model}
We present a theoretical model that accounts for these results
quantitatively. We start by considering a two-photon state that
represents both the spectral \emph{and} spatial characteristics of
photons, of the form
\begin{equation}\label{TwoPhotonState}
|\Psi\rangle=\int\int dxdx'\int\int d\Omega d\Omega'
\phi(x,x')\psi(\Omega,\Omega')|1_{x,\frac{\omega_{p}}{2}+\Omega},1_{x',\frac{\omega_{p}}{2}+\Omega'}\rangle,
\end{equation}
with the normalizations $\int\int dxdx'|\phi(x,x')|^{2}=1$ and
$\int\int d\Omega d\Omega'|\phi(\Omega,\Omega')|^{2}=1$. Here $x$
and $x'$ are spatial parameters for the signal and idler photons in
one dimension (the second dimension $y$ is not taken into
consideration without loss of generality); and $\Omega$ and
$\Omega'$ are deviations in frequency from the cental frequency
$\frac{\omega_{p}}{2}$. It is essential to note that the two-photon
state is assumed to be separable in the spatial--spectral
degrees-of-freedom. Thus, while the two photons may be entangled in
each of these degrees-of-freedom separately \cite{Barreiro06PRL},
there is no correlation between them. For a monochromatic pump with
large transverse spatial dimension (pump width larger than the
geometric mean of the pump wavelength and nonlinear-crystal
thickness \cite{Saleh00PRA}), the entangled state is correlated
spatially $\phi(x,x')=\phi(x)\delta(x-x')$ and anti-correlated
spectrally
$\psi(\Omega,\Omega')=\psi(\Omega)\delta(\Omega+\Omega')$. In this
formulation, the function $\psi(\Omega)$ is a baseband function
centered at frequency $\Omega=0$, and $\phi(x)$ is the transverse
spatial distribution of the pump \cite{Abouraddy07PRA}. The reduced
one-photon state obtained from Eq.~(\ref{TwoPhotonState}) is, in
general, mixed, and described by the density operator
\begin{equation}\label{OnePhotonState}
\rho=\int\int dxdx'\int\int d\Omega d\Omega'
\rho_{x}(x,x')\rho_{\Omega}(\Omega,\Omega')|1_{x,\frac{\omega_{p}}{2}+\Omega}\rangle\langle
1_{x',\frac{\omega_{p}}{2}+\Omega'}|,
\end{equation}
in which the spatial and spectral degrees-of-freedom remain
separable. For a monochromatic pump with large spatial dimension,
the state is characterized by
$\rho_{x}(x,x')=|\phi(x)|^{2}\delta(x-x')$ and
$\rho_{\Omega}(\Omega,\Omega')=|\psi(\Omega)|^{2}\delta(\Omega-\Omega')$.

Next consider the transformation to the state brought about by the
MZI and the MZIM. The MZI is characterized by two arms having the
same the number of reflections, imparted by mirrors or beam
splitters. It can be shown that the MZI conserves the separability
of the spatial and spectral degrees-of-freedom of the state of the
input optical field, such as the two-photon state set forth in
Eq.~(\ref{TwoPhotonState}) or the one-photon state provided in
Eq.~(\ref{OnePhotonState}). This can be shown by calculating the
fourth-order interference at the output,
$G^{(2)}(x_{1},x_{2};\tau)$. For the MZI, this fourth-order
coherence function is separable in the spatial and spectral
degrees-of-freedom:
$G^{(2)}(x_{1},x_{2};\tau)=G_{x}^{(2)}(x_{1},x_{2})G_{t}^{(2)}(\tau)$.
Since the detectors do not register the positions of the photons,
but instead integrate over the full transverse domain, the quantity
that is measured is $\int\int
dx_{1}dx_{2}G^{(2)}(x_{1},x_{2};\tau)$, which in this case is simply
$G_{t}^{(2)}(\tau)$. Assuming a monochromatic pump of large spatial
extent, we then have
\begin{equation}
G^{(2)}(\tau)=1-{\textstyle
\frac{1}{2}}\cos\omega_{p}\tau-{\textstyle \frac{1}{2}}\int
d\Omega|\psi(\Omega)|^{2}\cos2\Omega\tau,
\end{equation}
where we have assumed, for simplicity, that the spectral density
$|\psi(\Omega)|^{2}$ is an even function. The first term is a
constant background, the second is a sinusoid at the pump frequency,
and the third is the HOM dip.

The MZI also maintains the separability of the spatial and spectral
degrees-of-freedom of the one-photon state provided in
Eq.~(\ref{OnePhotonState}). The intensity at the output port thus
separates,
$G^{(1)}(x_{1},x_{1};\tau)=I(x_{1},\tau)=I_{x}(x_{1})I_{t}(\tau)$,
and the measured intensity as the delay is swept is simply $\int
dx_{1}I(x_{1},\tau)=I_{t}(\tau)$. This is given by
\begin{equation}\label{IntensityMZI}
I_{t}(\tau)=1-\cos\frac{\omega_{p}}{2}\tau \int
d\Omega|\psi(\Omega)|^{2}\cos\Omega\tau,
\end{equation}
which is the usual MZI interferogram. It is independent of the
spatial distribution and spatial coherence of the input field.

The consequences of removing a mirror from the MZI, which results in
an MZIM, are profound. The interferometer no longer preserves the
separability of the spatial and spectral degrees-of-freedom of the
input state of the field. When the number of mirrors in the two arms
of the MZI are not balanced, two copies of the input field reach the
detector, differing by a spatial inversion (in one spatial
dimension). Unless the field is an eigenfunction of the spatial
inversion process [such as even and odd functions $f(-x)=\pm f(x)$],
the two copies become distinguishable and do not interfere.

The fourth-order coherence function $G^{(2)}(x_{1},x_{2};\tau)$,
nevertheless, remains the same. Consider the case when the two
photons emerge together from the same port of the first beam
splitter in the interferometer (the case responsible for the
sinusoid at the pump frequency). The spatial probability amplitude
when the two photons travel through the arm without the spatial
flipper is $\phi(x,x')$, while that in the spatial-flipper arm is
$\phi(-x,-x')=\phi(-x)\delta(x-x')$. If the pump transverse spatial
distribution is even $\phi(-x)=\phi(x)$, which is the case in our
experiment, we have $\phi(-x,-x')=\phi(x,x')$, and the MZIM sinusoid
will be identical to that of the MZI. Note that there would result a
shift of $\pi$, but no change in amplitude, if the pump has an odd
spatial distribution $\phi(-x)=-\phi(x)$, and there would be a
reduction in amplitude if the pump has an arbitrary spatial
distribution (neither even nor odd). Consider now the case when the
two photons emerge from different ports of the first beam splitter
(the case responsible for the HOM dip). The two spatial probability
amplitudes interfering at the second beam splitter are
$\phi(x,-x')=\phi(x)\delta(x+x')$ and
$\phi(-x,x')=\phi(-x)\delta(x+x')$. One again, if the pump has an
even spatial distribution, the resulting HOM dip is identical to the
MZI case. As a result, the fourth-order coherence function is
identical for the MZI and MZIM as long as the pump has an even
spatial distribution, which is the case in our experiment.

Now consider the second-order coherence function $I(x_{1},\tau)$,
which differs from that in Eq.~(\ref{IntensityMZI}):
\begin{equation}
I_{t}(\tau)=1-|\alpha|\cos\left(\frac{\omega_{p}}{2}\tau-\varphi\right)
\int d\Omega|\psi(\Omega)|^{2}\cos\Omega\tau,
\end{equation}
where
\begin{equation}
\alpha=|\alpha|e^{i\varphi}=\int dx\rho_{x}(-x,x).
\end{equation}
The visibility of the temporal interferogram is thus found to be
weighted by a factor $\alpha$ that is a functional of the spatial
coherence of the source of the form $\int dx \rho_{x}(x,-x)$. A
field that is spatially incoherent, as in our experiments,
$\rho_{x}(x,x')=|\phi(x)|^{2}\delta(x-x')$ results in $\alpha=0$,
thus extinguishing the temporal interference. In other words, since
the one-photon field is \textit{spatially} incoherent, the field and
its spatially flipped version are mutually incoherent, resulting in
the loss of second-order \textit{temporal} interference.

\section{Conclusion}
In conclusion, we reported two similar experiments that share
identical fourth-order interference patterns, but whose second-order
interference behaviors are dramatically different. We explain our
results in terms of the spatial coherence properties of SPDC,
specifically that the two photons possess full spatial coherence
when considered jointly, but retain no spatial coherence when
considered individually. We remark that the MZIM used to observe
this effect may also be used for the measurement of arbitrary
optical fields, and thus offers us an important new tool for
exploring spatial coherence.

\section*{Acknowledgments}This work was supported by a U.~S.~Army
Research Office (ARO) Multidisciplinary University Research
Initiative (MURI) Grant and by the Center for Subsurface Sensing and
Imaging Systems (CenSSIS), an NSF Engineering Research Center. A.~F.~A. acknowledges the generous support and
encouragement of Y.~Fink and J.~D.~Joannopoulos.

\end{document}